\documentclass[11pt, reqno, oneside]{amsart}   	
\usepackage{amsaddr}
\usepackage{geometry}                		
\usepackage[parfill]{parskip}    		
\usepackage{graphicx}				
\usepackage{amssymb}
\usepackage{amsmath}
\usepackage{amsfonts}
\usepackage{mathbbol}
\usepackage{mathtools}
\usepackage{cite}

\usepackage[english]{babel}
\usepackage[utf8x]{inputenc}
\usepackage{color}

\newcommand{\T}{\mathbb{T}}

\newcommand{\mcW}{\mathcal{W}}
\newcommand{\mcV}{\mathcal{V}}
\newcommand{\mcO}{\mathcal{O}}

\newcommand{\mcA}{\mathcal{A}}

\newcommand{\mcH}{\mathcal{H}}
\newcommand{\mcG}{\mathcal{G}}
\newcommand{\N}{\mathrm{N}}
\newcommand{\M}{\mathrm{M}}
\newcommand{\Z}{\mathbb{Z}}

\newcommand{\g}{\mathrm{g}}

\newcommand{\mfB}{\mathfrak{B}}
\newcommand{\mfU}{\mathfrak{U}}

\newcommand{\mfP}{\mathfrak{P}}
\newcommand{\mfG}{\mathfrak{G}}
\newcommand{\mrZ}{\mathrm{Z}}

\title[]{The Variational Problem and Background Field in the Renormalization Group Method for Non-Linear Sigma Models}
\author{Abhishek Goswami}
\address{Faculty of Mathematics and Computer Science,
Adam Mickiewicz University in Pozna$\acute{\text{n}}$, 61-614 Pozna$\acute{\text{n}}$, Poland}
\curraddr{}
\email{abhishek.goswami@amu.edu.pl}
\date{\today}							

\begin{document}

\begin{abstract}
We study the variational problem as described by Balaban in his renormalization group method for Yang-Mills
theories in $d = 3, 4$ and adapt it to a class of Non-Linear Sigma Models in $d=2$. The result of the variational
problem is a minimal configuration which can serve as a classical background field in the renormalization group analysis. 
\end{abstract}
\maketitle

\section{Introduction}
Balaban with his renormalization group method has managed to establish the ultraviolet stability of the 
lattice Yang-Mills theories in $d=3, 4$ \cite{B_6, B_7, B_avg, B_prop, B_10, B_spaces, B_var, B_12, B_13, B_14, B_15, B_16}. 
His work in four dimensional Yang-Mills theories is 
described as a qualitative breakthrough \cite{JW}. However, Balaban's renormalization group method has remained
largely inaccessible to readers as the analysis is difficult to follow and is spread over several papers \cite{MRS}.
The goal of our program is to apply Balaban's renormalization group method to study the ultraviolet problem in
Euclidean Non-Linear Sigma Models in $d=2$. The Non-Linear Sigma Models in two dimensions are both 
critical and asymptotically free just like the Yang-Mills theories in four dimensions.

Here we focus on understanding the variational problem in the renormalization group method as discussed by Balaban
in \cite{B_var}. We study it for Non-Linear Sigma Models.
The result of the variational problem is a classical background field. The fluctuations about this background
field configuration denote the quantum corrections. 

The background field method is also widely used in physics especially while quantizing gauge theories, see for example, 
\cite{Ab, HPS} for applications. This is because the method preserves the symmetries of the theory while 
computing quantum corrections.

\subsection{Model}
We work on a lattice which is a two dimensional torus given by 
\begin{equation}
\label{1-14}
\T^{-\N}_{\M} = \big(L^{-\N}\Z / L^{\M}\Z \big)^{2} \hspace{1 cm} \text{Vol.} = L^{2 \M} 
 \hspace{1 cm} \text{lattice spacing} = L^{-\N}
\end{equation}
where $\M$ and $\N$ are large positive integers while $L$ is a large positive odd integer. 
The lattice notation we use here is due to Dimock \cite{D}.
From this torus we construct a (scaled) lattice with spacing $L^{-k}$ as $\T^{-k}_{\M+\N-k}$. We also consider a unit lattice
(the lattice of centers of unit blocks) given by
\begin{equation}
\label{-14}
\T^{0}_{\M+\N-k} = \T^{-k}_{\M+\N-k} \cap \Z^{2}
\end{equation}
and an $L$ lattice (the lattice of centers of $L \times L$ blocks) given by
\begin{equation}
\label{-14}
\T^{1}_{\M+\N-k} = \T^{-k}_{\M+\N-k} \cap (L\Z)^{2}
\end{equation}
Let $\eta = L^{-k}$ and denote $\T_{\eta} \equiv \T^{-k}_{\M+\N-k}$. Consider the field configurations U
\begin{center}
$\text{U} : \T_{\eta} \rightarrow \text{G}$ 
\end{center}
where G is a Lie group S\text{U}(2) in which case we have a O(4) Non-Linear Sigma Model. More generally
we take G as S\text{U}(n) and then we have a class of Non-Linear Sigma Models known as Non-Linear Chiral Models.

Next we rescale to the unit lattice $\T^{0}_{\M+\N}$. The field configurations
$\text{U}_{L^{k}}: \T^{0}_{\M+\N} \rightarrow \text{G}$ are given by 
\begin{equation}
\label{1-19}
\text{U}_{L^{k}}(L^{k}x) = \text{U}(x) \hspace{1 cm} x \in \T_{\eta}
\end{equation}
As the field configurations take values on the group manifold and not on the lattice, rescaling of the lattice
has no effect on them. This property of the field configurations to be constrained on the group manifold is what makes
the model non-linear.

Let $\langle x, x + \eta e_{\mu}\rangle$ and $\langle x, x - \eta e_{\mu}\rangle$ denote oriented bonds starting from $x$ 
on the lattice $\T_{\eta}$. Define  
\begin{equation}
\label{-14}
\partial \text{U} (x, x+\eta e_{\mu}) = \text{U}(x) \text{U}^{-1}(x+\eta e_{\mu}) \hspace{1 cm}
\partial \text{U} (x, x-\eta e_{\mu}) = \text{U}(x) \text{U}^{-1}(x-\eta e_{\mu})
\end{equation} 
\underline{\textit{Action}}. We work with Wilson-like action as remarked by Balaban 
(Eq. 0.32 in \cite{B_12}) for Non-Linear Chiral Models.
Let $b$ be an oriented bond between two sites $x$ and $y$ on $\T_{\eta}$. 
Denote the boundary of $b$ as $\partial b = \langle x y \rangle$. The action functional is
\begin{equation}
\label{4-14}
\mcA(\text{U}) = \sum_{\partial b \subset \T_{\eta}} \eta^{d-2} [1 - \text{Re Tr}\hspace{0.05 cm} \text{U}(\partial b)]
\end{equation}
with $d=2$ where
\begin{equation}
\label{-14}
\text{U}(\partial b) = (\partial \text{U})(b) = \text{U}(x) \text{U}^{-1}(y)
\end{equation}
Note that as $d=2$ the functional ($\ref{4-14}$) is independent of the lattice spacing.

\underline{\textit{Symmetries of $\mcA$}}. First rewrite $\mcA$ as
\begin{equation}
\label{7-14}
\mcA(\text{U}) = \sum_{\langle x y \rangle \subset \T_{\eta}} [1 - \text{Re Tr}\hspace{0.05 cm} \{\text{U}(x) \text{U}^{-1}(y)\}]
\end{equation}
Let $v, w \in $ SU(n). Consider a global transformation $\text{U} \rightarrow v \text{U} w$. Then
\begin{equation}
\label{-14}
\begin{aligned}
\mcA(v \text{U} w) &= \sum_{\langle x y \rangle \subset \T_{\eta}} 
[1 - \text{Re Tr}\hspace{0.05 cm} \{ v\text{U}(x)w  w^{-1}\text{U}^{-1}(y)v^{-1}\}] \\
&= \sum_{\langle x y \rangle \subset \T_{\eta}} 
[1 - \text{Re Tr}\hspace{0.05 cm} \{v^{-1} v\text{U}(x) \text{U}^{-1}(y)\}] = \mcA(\text{U})
\end{aligned}
\end{equation}
where we have used the cyclic property of the trace.

We are interested in the integrals of the form
\begin{equation}
\label{-14}
\mrZ = \int d\text{U} \hspace{0.1 cm} e^{-\frac{1}{g^{2}} \mcA(\text{U}) - E }
\end{equation}
where $g$ is the coupling constant which is dimensionless in $d=2$, 
$E$ is the vacuum energy and $d\text{U} = \prod_{x \in \T_{\eta}} d\text{U}(x)$ is the 
product Haar measure of the group G. The integral is over the group manifold. We intend to study such integrals using the method 
of steepest descents. This is how the variational problem enters into the analysis. To do so we will have to define
the integral over a linear vector space instead of the group manifold. We do that by transforming to the local 
coordinates of U given by Lie algebra variables; $A$, as $\text{U} = e^{i A}$ . 
First we give an overview of the method of steepest descents as applied in our analysis.

\subsection{Steepest descent}

This discussion follows the one in \cite{BJ}. 
Let $A$ be an element of some finite domain in a linear vector space. Consider the integral
\begin{equation}
\label{-14}
I = \int dA \hspace{0.1 cm} e^{-\big[\frac{1}{g^{2}} f(A) + E(g, A) \big]}
\end{equation}
where $g$ is finite and $E$ is a regular function of $g$ and is sufficiently small.
We assume that $f$ has a unique critical point in the domain which is a minima. Denote the minima by $A_{0}$.
We look to expand $f$ about $A_{0}$ and hence $A_{0}$ is called a background. 
Let $A - A_{0} = g A^{\prime}$ denote the scaled fluctuation variable. Then the integral $I$ becomes
\begin{equation}
\label{11-14}
I = g^{\text{dim}} \hspace{0.05 cm} e^{-\frac{1}{g^{2}} f(A_{0}) - E(g, A_{0})} \int 
e^{-\frac{1}{2} \langle A^{\prime}, d^{2}(A_{0})f A^{\prime} \rangle + V} d A^{\prime} 
\end{equation}
Here $V$ is a function that includes higher order $\mcO({A^{\prime}}^{3})$ terms in the expansion of $f$ and 
$d^{2}(A_{0})f$ denotes the quadratic form also known as Hessian of $f$. 
The integral in $(\ref{11-14})$ is then studied using cluster expansion. The output of the cluster expansion
is a function of the background $A_{0}$ which is regular in $g$ and also sufficiently small.

Our goal here is to understand the background configuration which is the minimum of the functional 
as written above. The cluster expansion and analysis of the coupling constant will be carried out in the future work.

Before we can formally state the variational problem we need few more definitions.

\subsection{Block averaging}
Consider the $L^{-k}$ lattice $\T_{\eta}$ and a $L^{-k+1}$ lattice given by $\T^{-k+1}_{\M+\N-k}$. 
For a site $y \in \T^{-k+1}_{\M+\N-k}$ define a block centered on $y$ as
\begin{equation}
\label{-14}
B(y) = \Big\{x \in \T_{\eta} : \sup_{\mu}|x_{\mu} - y_{\mu}| \leqslant L^{-k+1}/2\Big\}
\end{equation}
Note that since $L$ is odd the boundary of the block lies between the two neighboring sites.

We define a one step block averaging operation that maps the field configurations on $\T^{-k}_{\M+\N-k}$ to the ones on 
$\T^{-k+1}_{\M+\N-k}$ as 
\begin{equation}
\label{13-14}
\bar{\text{U}}(y) = \text{exp} \hspace{0.05 cm} \Big[\sum_{x \in B(y)} \frac{1}{L^{2}} 
\text{log}\hspace{0.05 cm}\big(\text{U}(x) \text{U}^{-1}(y)\big)\Big] \text{U}(y)
\end{equation}
Note that if $\text{U}(x) = e^{i A(x)}$ and $\text{U}(y) = e^{i A(y)}$ with $A$ small then using
Baker-Campbell-Hausdorff formula: 
$\text{log}\hspace{0.05 cm} [e^{i A(x)} e^{-i A(y)}] = i A(x) - i A(y) + \frac{1}{2}[A(x), A(y)] + \cdots$, we get
\begin{equation}
\label{-14}
\bar{\text{U}}(y) = \text{exp} \hspace{0.05 cm} \Big[i \bar{A} + \mcO(A^{2}) \Big], \hspace{1 cm} 
\bar{A} = \sum_{x \in B(y)} \frac{1}{L^{2}} A(x) 
\end{equation}
Let $v, w \in $ SU(n). Then under a global transformation $\text{U} \rightarrow v\text{U}w$, first note that
\begin{center}
$v\text{U}(x)w [v\text{U}(y)w]^{-1} = v\text{U}(x)\text{U}^{-1}(y)v^{-1}$
\end{center}
has same eigenvalues as $\text{U}(x) \text{U}^{-1}(y)$ and hence is unitary equivalent to $\text{U}(x) \text{U}^{-1}(y)$. 
Then by the spectral orthogonal projection representation (Eq. 22 in \cite{B_avg}), their logarithms are also unitary
equivalent
\begin{center}
$\text{log}\hspace{0.05 cm}\big(v\text{U}(x)\text{U}^{-1}(y)v^{-1}\big) = 
v \hspace{0.05 cm} \text{log}\hspace{0.05 cm}\big(\text{U}(x)\text{U}^{-1}(y)\big) v^{-1} $
\end{center}
Thus, we have
\begin{equation}
\label{15-14}
\begin{aligned}
\overline{v\text{U}w} &= \text{exp} \hspace{0.05 cm} \Big[\sum_{x \in B(y)} \frac{1}{L^{2}} 
\text{log}\hspace{0.05 cm}\big(v\text{U}(x)w w^{-1}\text{U}^{-1}(y)v^{-1}\big)\Big] v\text{U}(y)w \\
&= \text{exp} \hspace{0.05 cm} \Big[v \sum_{x \in B(y)} \frac{1}{L^{2}} 
\text{log}\hspace{0.05 cm}\big(\text{U}(x) \text{U}^{-1}(y)\big)v^{-1}\Big] v\text{U}(y)w \\
&= v \bar{\text{U}} w
\end{aligned}
\end{equation}
the last step follows from the taylor expansion of the exponential function since
for a matrix $X$, $(v X v^{-1})^{n} = v X^{n} v^{-1}$.

A composition of averaging operations $k$ times is same as averaging over a block of side length $L^{k}$ and volume
$L^{2k}$. Let $y \in \T^{0}_{\M+\N-k}$. Define a block centered on $y$
\begin{equation}
\label{-14}
B^{k}(y) = \Big\{x \in \T^{-k}_{\M+\N-k} : \sup_{\mu}|x_{\mu} - y_{\mu}| \leqslant \frac{L^{k}}{2}\Big\} 
\end{equation}
$B^{k}(y)$ is a block with $L^{k}$ sites on each side. Then
\begin{equation}
\label{-14}
\bar{\text{U}}^{(k)} : \text{U} \hspace{0.2 cm} \text{on} \hspace{0.2 cm} \T^{-k}_{\M+\N-k} \rightarrow 
\text{U}_{L^{k}} \hspace{0.2 cm} \text{on} \hspace{0.2 cm} \T^{0}_{\M+\N-k}
\end{equation}

\subsection{Spaces of regular configurations} 
Let $M = L^{m}$ be a very large positive integer ($M$ is not same as the vol. parameter in ($\ref{1-14}$)).
Then the lattice of the centers of $M \times M$ blocks given by
\begin{center}
$\T^{m}_{\M+\N-k} = \T^{-k}_{\M+\N-k} \cap (M\Z)^{2}$
\end{center}
partitions the $\T_{\eta}$ lattice into a sum of $M$ blocks. 
We call a union of $M$ blocks as $\Omega_{k}$. 

We take $\Omega_{k} =\T_{\eta}$.
Let $\text{U}$ be a field configuration on the lattice $\T_{\eta}$. Let $\varepsilon_{0} > 0$. Define a space 
\begin{center}
$\mfU_{k}(\varepsilon_{0}, \Omega_{k}) = \mfU_{k}(\varepsilon_{0})$
\end{center}
of the field configurations by the conditions
\begin{equation}
\label{19-14}
\begin{aligned}
|(\partial \text{U})(b) - 1| &< \varepsilon_{0}L^{-k} = \varepsilon_{0}\eta , \hspace{0.5 cm} b \in \Omega_{k}
  \\
|\delta\text{U}(x)| &< \varepsilon_{0}L^{-k} = \varepsilon_{0}\eta, \hspace{0.5 cm} x \in \Omega_{k}
\end{aligned}
\end{equation}
where
\begin{equation}
\label{-19}
\delta\text{U}(x) = \sum_{\mu} \big[ \text{U}(x- e_{\mu})\text{U}(x)^{-1} - \text{U}(x)\text{U}(x+ e_{\mu})^{-1} \big]
\end{equation}
The norms are operator norms for a matrix $X$, \lq\lq the operator norm $|X| = \sup_{|\psi| = 1} |X\psi|"$.
A field configuration U belonging to $\mfU_{k}(\varepsilon_{0})$ is called regular.

Let $\varepsilon_{1} < \varepsilon_{0}$. 
Define a space $\mfB_{k}(\varepsilon_{1})$ as
\begin{equation}
\label{21-14}
\mfB_{k}(\varepsilon_{1}) = 
\{\max_{b} |\partial \text{V}(b) - 1| < \varepsilon_{1} \hspace{0.1 cm} \text{for} \hspace{0.1 cm} b \in \Omega_{k} \cap \Z^{2} \}
\end{equation}
Define another space $\mfB_{k, 1}(\varepsilon_{1})$ as
\begin{equation}
\label{22-14}
\mfB_{k, 1}(\varepsilon_{1}) = 
\{\max_{b} |\partial \text{W}(b) - 1| < \varepsilon_{1} \hspace{0.1 cm} \text{for} \hspace{0.1 cm} b \in \Omega_{k} \cap (L\Z)^{2} \}
\end{equation}

\textbf{Balaban's notation}. On the lattice $\T_{\eta}$ Balaban \cite{B_var} considers a sequence of domains $\{\Omega_{j}\}$ for
 $j=0, 1,\cdots, k$ with $\T_{\eta} \supset \Omega_{0} \supset \Omega_{1} \supset \cdots \supset \Omega_{k}$ such that
  \begin{itemize}
  \item $\Omega_{j}$ is a union of $M (L^{j}\eta)$ blocks.
  \item For any domain $\Omega \subset \T_{\eta}$, $\Omega^{(j)} = \Omega \cap (L^{j}\eta \Z)^{2}$. Thus, $\Omega^{(j)}$
  are centers of $L^{j}$ blocks in $\Omega$ which are points on the lattice $\T^{-(k-j)}_{\M+\N-k}$.
\end{itemize}
The sequence admits the case where some domains are equal to the lattice $\T_{\eta}$.
Balaban also considers the sets of the form $\Omega_{j}^{(j)}/\Omega_{j+1}^{(j)}$ for $j=0, 1,\cdots, k-1$
for spaces $(\ref{21-14}), (\ref{22-14})$. Here we work with the case where all $\Omega_{j}$ equal to the
entire lattice $\T_{\eta}$. Hence, we work with a single scale determined by $\Omega_{k}$ 
for spaces $(\ref{21-14}), (\ref{22-14})$. This is sufficient for the first treatment of the model as we work in a
small field approximation. For the full model we have to consider the multiscale geometry.

Note that on the scale determined by $j$, where a bond variable as in $(\ref{19-14})$ has a bond length of  
$L^{j}\eta$, we have $|(\partial \text{U})(b) - 1| < \varepsilon_{0} \eta (L^{j}\eta)^{-1}$ for $b \in \Omega_{j}$.

\subsubsection{Symmetry subgroup} Denote $\mfB_{k} \equiv \T^{0}_{\M+\N-k}$. 
Let V be a fixed configuration on $\mfB_{k}(\varepsilon_{1})$. 
Define a space $\mfB_{k}(\mfB_{k}, \text{V})$ for configurations U  satisfying
\begin{equation}
\label{1-15}
\bar{\text{U}}^{(k)} = \text{V}  \hspace{0.2 cm} \text{on} \hspace{0.2 cm} \mfB_{k}
\end{equation}
Since V is fixed, any global transformation must keep $(\ref{1-15})$ invariant and
hence,
\begin{equation}
\label{2-15}
u \text{V} w = \text{V} \hspace{0.2 cm} \text{on} \hspace{0.2 cm} \mfB_{k}
\end{equation}
\textbf{Remark 1}. As any global transformation is independent of the lattice points, note that if V is \textit{any constant configuration} 
and $u$ is \textit{any} element of SU(n) then $w = \text{V}^{-1} u^{-1} \text{V}$ imply ($\ref{2-15}$) as $u \text{V} w = \text{V}$.

When V is not a constant configuration, the transformations that satisfy $(\ref{2-15})$ are of the form $u\text{V}u^{-1}$.
Such transformations commute with $\text{V}$; $[u, \text{V}] = 0$ and 
thus, $u \text{V} = \text{V} u$ imply $u \text{V} u^{-1} = \text{V}$. Denote the subgroup formed by the transformations 
satisfying $(\ref{2-15})$ as
\begin{equation}
\label{25-21}
\mcG(\text{V}) = \big\{u \in \text{SU(n)} \hspace{0.2 cm} \big| \hspace{0.2 cm} 
u \text{V} u^{-1} = \text{V}  \hspace{0.2 cm} \text{on} \hspace{0.2 cm} \mfB_{k} \big\}
\end{equation} 
 
For any group G = SU(n), the only element that commutes with every other element is the identity.
Thus, for a general V, the subgroup $\mcG(\text{V})$ is just the identity.

We can similarly define a space $\mfB_{k, 1}(\mfB_{k, 1}, \text{W})$ for configurations U satisfying
 $(\ref{1-15})$ but for $(k+1)$ instead of $(k)$ with W instead of V and on  $\mfB_{k, 1}$.

\subsubsection{New space} Consider a set of two new field configurations $\text{U}_{0}$ and $\text{U} = \text{U}^{\prime} \text{U}_{0}$. 
Here we define a space whose arbitrary configuration is of the form $\text{U}^{\prime} \text{U}_{0}$.
Let $a_{0}, a_{1} > 0$. We assume that
\begin{equation}
\label{3-15}
\begin{aligned}
&\text{U}_{0} \in \mfU_{k}(a_{0}), \hspace{0.2 cm} \text{U}^{\prime} \text{U}_{0} \in \mfU_{k}(a_{0}) \cap \mfB_{k}(\mfB_{k}, \text{V}) \\
& |(\overline{\text{U}^{\prime} \text{U}_{0}})^{(k)} -  \bar{\text{U}}_{0}^{(k)}| < a_{1} 
\hspace{0.2 cm} \text{on} \hspace{0.2 cm} \mfB_{k}
\end{aligned}
\end{equation}
Note that by $(\ref{1-15})$ and the first line in $(\ref{3-15})$, 
$(\overline{\text{U}^{\prime} \text{U}_{0}})^{(k)} = \text{V}$. Then the second line in $(\ref{3-15})$
is satisfied when V is close to $\bar{\text{U}}_{0}^{(k)}$ that is $|\text{V} - \bar{\text{U}}_{0}^{(k)}| < a_{1}$.

\subsection{Variational problem statement}
Let $\mcA^{(k)}$ denote the functional $\mcA(\text{U})$ defined on $\T^{-k}_{\M+\N-k}$.
Then the variational problem is defined as
\begin{equation}
\label{-14}
\boxed{\text{U} \rightarrow \mcA^{(k)}(\text{U}), \hspace{0.5 cm} 
\text{U}  \in \mfU_{k}(\varepsilon_{0}) \cap \mfB_{k}(\mfB_{k}, \text{V}), \hspace{0.5 cm} \bar{\text{U}}^{(k)} = \text{V}}
\end{equation}

\textbf{Theorem 1} Let $a_{0}, a_{1}$ be strictly positive constants. Then for 
$0 < \varepsilon_{0} \leqslant a_{0}$ and $0 < \varepsilon_{1} \leqslant a_{1}$ let $\text{V} \in \mfB_{k}(\varepsilon_{1})$ be fixed. 
There exists a constant $B_{3} > 0$ obeying $B_{3} \varepsilon_{1} \leqslant \varepsilon_{0}$ such that
a unique minimal configuration of the functional $\mcA^{(k)}(\text{U})$ exists in the space
\begin{center}
$\big\{\text{U} \in \mfU_{k}(B_{3}\varepsilon_{1}) \hspace{0.2 cm} \big| \hspace{0.2 cm} 
u \text{U} u^{-1} = \text{U}  \hspace{0.2 cm} \text{for all} \hspace{0.2 cm} u \in \mcG(\text{V}) \big\} \cap \mfB_{k}(\mfB_{k}, \text{V})$. 
\end{center}

\textbf{Remark 2} In the $k^{\text{th}}$ step of the actual renormalization group construction, the action used is the
effective action for scale $k$, which includes, in addition to the $\mcA^{(k)}$ of Theorem 1, both contributions from the
previous fluctuation integrals and the effects of renormalization. 
$\mcA^{(k)}$ is only the classical contribution to the effective action and hence
the background configuration is classical as well. Note that only the pure small fields, i.e. $\text{U}$ with 
$|\partial\text{U} - 1|$ sufficiently small, are being considered in this paper. In the actual renormalization group
construction all $\text{U}'s$ must be considered.

\underline{\textit{Outline}} In Sect. $\ref{RG}$ we first understand the above statement of the variational problem in the 
context of the renormalization group (RG) transformation. Then the proof of the Theorem 1 follows in the Sect. $\ref{proof}$.
In the final Sect. $\ref{Analyticity}$ we discuss the analyticity of the background configuration.

\section{RG transformation}\label{RG}

Here we give a brief overview of the renormalization group (RG) transformation with the aim to understand 
the variational problem and the role of critical configuration in a single iteration. RG iteration has three steps: 
integration, rescaling and extension.

Consider the functional $\mcA(\text{U})$ on the unit lattice $\T^{0}_{\M+\N}$ with
$\text{U} \in \mfU_{0}(\varepsilon_{0})$. Let $\text{V} \in \mfB_{0,1}(\varepsilon_{1})$.
Then the transformation is defined as \cite{BJ}
\begin{equation}
\label{-14}
(T\rho)(\text{V}) = \int d\text{U} \hspace{0.05 cm} \delta(\bar{\text{U}}\text{V}^{-1}) 
\hspace{0.05 cm} e^{-\frac{1}{g^{2}} \mcA(\text{U}) - E }
\end{equation}
The delta function introduces the constraint $\bar{\text{U}} = \text{V}$. Let $v, w \in$ SU(N). Under a global
transformation $\text{U} \rightarrow v \text{U} w$ and $\text{V} \rightarrow v \text{V} w$, we have seen that
in $(\ref{15-14})$, $\overline{v\text{U}w} = v\bar{\text{U}}w$. Hence 
$\delta(\overline{v\text{U}w} w^{-1}\text{V}^{-1}v^{-1}) =  \delta(v\bar{\text{U}}\text{V}^{-1}v^{-1}) = \delta(\bar{\text{U}}\text{V}^{-1})$
due to the invariance of the delta function at the identity of the group, $\delta(v I v^{-1}) = \delta(I)$. 
Since the Haar measure $d$U also remains invariant we have $(T\rho)(v\text{V}w) = (T\rho)(\text{V})$.

The first step introduces a split between the
short distance (or high momenta) degrees of freedom and long distance (or low momenta) degrees of freedom.
The unit lattice configurations here correspond to the high momenta degrees of freedom
while the $L$ lattice configurations correspond to the low momenta degrees of freedom.

We analyze the above integral by the method of steepest descent. This is where we require a minimal
configuration of the functional  $\mcA(\text{U})$ with delta function constraint. The minimal configuration
is the high momenta degree of freedom which serves as a background field. 
We integrate out the fluctuation about this high momenta background and then scale back to the unit lattice for convenience.

This procedure is carried out until we reach a unit scale denoted as $L^{k}\epsilon = 1$, where $\epsilon = L^{-K}$.
Since we are rescaling back to the unit lattice after each integration, we are working with
$L^{-k}, \cdots, 1, L$ lattices. 

Note that the unit lattice at the start of the renormalization is 
$\mathbb{T}_{\mathrm{M+N}}^{0}$. Then the unit lattice at the end of $k$ RG iterations is
\begin{center}
$\mathbb{T}_{\mathrm{M+N}-k}^{0} = \mathbb{T}_{\mathrm{M+N}-k}^{-k} \cap \mathbb{Z}^{2} $
\end{center}
and a $L-$ lattice during the $(k+1)^{\text{th}}$ iteration is
\begin{center}
$\mathbb{T}_{\mathrm{M+N}-k}^{1} = \mathbb{T}_{\mathrm{M+N}-k}^{-k} \cap (L \mathbb{Z})^{2}$.
\end{center}
Rescaling corresponds to $\mathbb{T}_{\mathrm{M+N}-k}^{1} \rightarrow \mathbb{T}_{\mathrm{M+N}-k-1}^{0}$.
For our model, rescaling does not introduce any canonical dimension dependent changes to the field configurations
as shown in $(\ref{1-19})$. 

The last step extension will be mentioned shortly.

\subsection{Defining the variational problem}

We first discuss the role of a critical configuration in a RG iteration. After that we define the variational problem in a manner 
such that its solution \lq\lq that is the existence of a unique critical configuration" would mean that the RG iteration 
can be started.

\underline{$k=1$} At the start of the first iteration we have
\begin{equation}
\label{ }
\mcA(\text{U}), \hspace{0.5 cm} \text{U}  \in \mfU_{0}(\varepsilon_{0}) \cap \mfB_{0, 1}(\mfB_{0, 1}, \mcV),
\hspace{0.5 cm} \bar{\text{U}} = \mcV \nonumber
\end{equation}
The action functional $\mcA(\text{U})$ is functional of only one variable U. The
critical configuration $\text{U}_{0}(\mcV)$ belongs to the space 
$ \mfU_{0}(B_{3}\varepsilon_{1})  \cap \mfB_{0, 1}(\mfB_{0, 1}, \mcV)$.
We take critical configuration $\text{U}_{0}(\mcV)$ as a background configuration and
expand the functional $\mcA(\text{U})$ around it. Note that as $\text{U}_{0}(\mcV)$ is defined on a unit lattice, its
fluctuation is also defined on a unit lattice which we integrate out. This completes the first step of the iteration.

Next we rescale back to the unit lattice so that
\begin{equation}
\text{U}_{0}(\mcV) \in \mfU_{0}(B_{3}\varepsilon_{1})  \cap \mfB_{0, 1}(\mfB_{0, 1}, \mcV)  \xrightarrow{\text{rescale}} 
\text{U}_{1}(\text{V}) \in \mfU_{1}(B_{3}\varepsilon_{1}) \cap  \mfB_{1}(\mfB_{1}, \text{V})  
\end{equation}

The final step extension corresponds to simply recognizing that $\text{U}_{1}(\text{V})$ is already in the space
$\mfU_{1} (\varepsilon_{0}) \cap \mfB_{1}(\mfB_{1}, \text{V})$ since $B_{3}\varepsilon_{1} \leqslant \varepsilon_{0}$.

At the end of the first iteration the functional of two variables U and V is written as
$\mcA^{(1)}(\text{U}_{1}(\text{V}))$. It is the minimal value of the functional $\mcA^{(1)}$ in variable U.
Thus, at the end of the first iteration the variational problem is defined by
\begin{equation}
\label{ }
\boxed{\text{U} \rightarrow \mcA^{(1)}(\text{U}), \hspace{0.5 cm} 
\text{U}  \in \mfU_{1}(\varepsilon_{0}) \cap  \mfB_{1}(\mfB_{1}, \text{V}), \hspace{0.5 cm} \bar{\text{U}} = \text{V}}
\end{equation}
which has a critical configuration $\text{U}_{1}(\text{V})  \in \mfU_{1}(B_{3}\varepsilon_{1}) \cap  \mfB_{1}(\mfB_{1}, \text{V})$.
Thus, showing the existence of $\text{U}_{1}(\text{V})$ implies that the first iteration can be started as described above.
It provides the background configuration $\text{U}_{0}(\mcV)$.

\underline{$k=2$} At the start of the second iteration we have
\begin{equation}
\label{ }
\mcA^{(1)}(\text{U}_{1}(\text{V})), \hspace{0.5 cm}
\text{V} \in \mfB_{1}, \hspace{0.5 cm} \mcW \in \mfB_{1,1}, \hspace{0.5 cm} \bar{\text{V}} = \mcW
\nonumber
\end{equation}
A critical configuration of the functional $\mcA^{(1)}(\text{U}_{1}(\text{V}))$ is
critical in both the variables U and V. We already have configuration $\text{U}_{1}(\text{V})$ which is critical in variable U.
Denote a critical configuration in V of the functional $\mcA^{(1)}(\text{U}_{1}(\text{V}))$ with constraint $ \bar{\text{V}} = \mcW$
as $\text{V}^{(1)}(\mcW) \in \mfB_{1}$. Then the critical configuration in both the
variables is $\text{U}_{1}(\text{V}^{(1)}(\mcW))$ belonging to the space
$\mfU_{1} (B_{3}\varepsilon_{1}) \cap \mfB_{1, 1}(\mfB_{1, 1}, \mcW)$. This serves as the background configuration.

After integrating out the fluctuation, we do the rescaling so that
\begin{equation}
\text{U}_{1}(\text{V}^{(1)}(\mcW)) \in \mfU_{1} (B_{3}\varepsilon_{1}) \cap \mfB_{1, 1}(\mfB_{1, 1}, \mcW)
\xrightarrow{\text{rescale}} \text{U}_{2}(\text{W}) \in \mfU_{2} (B_{3}\varepsilon_{1}) \cap \mfB_{2}(\mfB_{2}, \text{W})
\end{equation}
and
\begin{equation}
\label{3-13 }
\bar{\text{U}}_{2}^{(2)} = \text{W} \hspace{0.2 cm} \text{on} \hspace{0.2 cm} \mfB_{2} \hspace{1 cm}
\bar{\text{U}}_{2} = \text{V}^{(1)}  \hspace{0.2 cm} \text{on} \hspace{0.2 cm} \mfB_{1}
\end{equation}

Extension corresponds to simply recognizing that the background configuration $\text{U}_{2}(\text{W})$ 
is already in the space $\mfU_{2} (\varepsilon_{0}) \cap \mfB_{2}(\mfB_{2}, \text{W})$ 
since $B_{3}\varepsilon_{1} \leqslant \varepsilon_{0}$.

Thus, at the end of the second iteration the functional is $\mcA^{(2)}(\text{U}_{2}(\text{W}))$ and
the variational problem is defined by
\begin{equation}
\label{ }
\boxed{\text{U} \rightarrow \mcA^{(2)}(\text{U}), \hspace{0.5 cm} 
\text{U}  \in \mfU_{2}(\varepsilon_{0}) \cap \mfB_{2}(\mfB_{2}, \text{W}), \hspace{0.5 cm} \bar{\text{U}}^{(2)} = \text{W}}
\end{equation}
which has a critical configuration $\text{U}_{2}(\text{W}) \in \mfU_{2} (B_{3}\varepsilon_{1}) \cap \mfB_{2}(\mfB_{2}, \text{W})$.
Thus, once again showing the existence of a unique critical configuration $\text{U}_{2}(\text{W})$ would imply that
the second RG iteration can be started as described above.

\underline{After $k$ iterations} We have a functional of two variables U and V as $\mcA^{(k)}(\text{U}_{k}(\text{V}))$.
The variational problem is defined as 
\begin{equation}
\label{34-21}
\boxed{\text{U} \rightarrow \mcA^{(k)}(\text{U}), \hspace{0.5 cm} \text{U}  \in \mfU_{k}(\varepsilon_{0}) \cap 
\mfB_{k}(\mfB_{k}, \text{V}), \hspace{0.5 cm} \bar{\text{U}}^{(k)} = \text{V}}
\end{equation}
which has a critical configuration $\text{U}_{k}(\text{V}) \in \mfU_{k} (B_{3}\varepsilon_{1}) \cap \mfB_{k}(\mfB_{k}, \text{V})$.

\section{Proof}\label{proof}

We start by recalling definitions of some operators from \cite{B_prop}. After that we discuss
a mapping transformation from \cite{B_spaces}. We use the mapping transformation to reduce the proof of Theorem 1
to proving a Proposition 4. We then first give a detailed proof of the Proposition 4 in Subsect. $\ref{expansion}$ using 
perturbative methods. Proof of Theorem 1 follows in the next Subsect. $\ref{proof1}$.

\subsection{Minimizers}
The minimizer is a solution of the linear variational problem defined as (Eqs. 3.109, 3.110 \cite{B_prop})
\begin{equation}
\label{-14}
A \rightarrow \frac{1}{2} \langle A, \Delta A\rangle;  \hspace{1 cm} QA =B
\end{equation}
where the constraint $Q$ is linear. Both $Q$ and $\Delta$ depend on the background. 
The minimizer is a linear operator $H$ and has a representation (Eq. 1.103 in \cite{B_6} or Eq. 3.126 \cite{B_prop})
\begin{equation}
\label{30-14}
H B = G Q^{\ast}(QGQ^{\ast})^{-1}B
\end{equation}
where $G = (\Delta + Q^{\ast}a Q)^{-1}$, $a$ is a positive constant. $G$ has a kernel that decays exponentially,
see for example Appendix D in \cite{D} (with background as identity).We consider another minimizer by replacing 
the quadratic form $\Delta$ in the linear variational problem as (Eq. 3.127 \cite{B_prop})
\begin{equation}
\label{-14}
A \rightarrow \frac{1}{2} \langle A, \Delta A \rangle - \langle H C^{(2)}(A), \text{J} \rangle; \hspace{1 cm}
QA =B
\end{equation}
where $C^{(2)}(A)$ is a quadratic polynomial in $A$ which takes values on the unit lattice and J is defined by $(\ref{43-14})$. 
The minimizer $H_{1}$ has a representation (Eq. 3.129 \cite{B_prop})
\begin{equation}
\label{32-14}
H_{1} B = G_{1} Q^{\ast}(QG_{1}Q^{\ast})^{-1}B
\end{equation}
where $G_{1} = (\Delta -H^{\ast}\text{J} + Q^{\ast}a Q)^{-1}$.

Since both the minimizers $H$ and $H_{1}$ have representations in terms of the Green's function, they both have
exponentially decaying kernels. They also satisfy the bounds (Eqs. 46, 103  \cite{B_var})
\begin{equation}
\label{33-14}
|H B| \leqslant B_{0}  |B|, \hspace{0.5 cm} |H_{1} B| \leqslant B_{1} |B| \hspace{0.3 cm} \text{on} \hspace{0.3 cm} \Omega_{k}
\end{equation}

\textbf{Remark 3} Minimizers and other operators appearing in our analysis have simpler representation
and properties than the similar operators in Balaban's Yang Mills analysis. This is because Yang Mills theories have 
an additional constraint due to the gauge fixing which further complicates their regularity properties.

\subsection{Mapping transformation} We denote the global transformations satisfying $(\ref{2-15})$ as 
$\text{U} \rightarrow \text{U}^{u}$
given by
\begin{equation}
\label{1-16}
\text{U}^{u} = u \text{U} u^{-1}
\end{equation}
We are interested in the configurations $\text{U}_{0}$ and $\text{U} = \text{U}^{\prime} \text{U}_{0}$ defined by $(\ref{3-15})$.
For a configuration $\text{U} = \text{U}^{\prime} \text{U}_{0}$, define a new symmetry transformation law 
keeping the configuration $\text{U}_{0}$ fixed as (Eq. 1.17, \cite{B_spaces}). Then for $\text{U}^{\prime}$ redefine 
($\ref{1-16}$) as
\begin{equation}
\label{1-18}
{\text{U}^{\prime}}^{u} = u \text{U}^{\prime} \text{U}_{0} u^{-1} \text{U}_{0}^{-1} 
\equiv u \text{U}^{\prime} R(\text{U}_{0}) u^{-1}
\end{equation}
Thus,
\begin{equation}
\label{2-18}
\text{U}^{u} =  u \text{U} u^{-1} = u \text{U}^{\prime} \text{U}_{0} u^{-1} = 
u \text{U}^{\prime} \text{U}_{0} u^{-1}\text{U}_{0}^{-1}\text{U}_{0} = {\text{U}^{\prime}}^{u} \text{U}_{0}
\end{equation}
Denote $R_{0} = R(\text{U}_{0})$. Recall that we are working with global transformations satisfying $(\ref{2-15})$. 
We replace this condition by the following condition (Eq. 1.29, \cite{B_spaces})
\begin{equation}
\label{3-18}
(\overline{R_{0}u}^{(k)})(y) = 1 \hspace{0.5 cm} y \in \mfB_{k}
\end{equation}
where $(\overline{R_{0}u}^{(k)})(y)$ is given by 
\begin{equation}
\label{4-18}
(\overline{R_{0}u}^{(k)})(y) = u \hspace{0.05 cm} \text{exp} \hspace{0.05 cm} \Big[\sum_{x \in B^{k}(y)} \frac{1}{L^{2}} 
\text{log}\hspace{0.05 cm} u \text{U}_{0}(x) u^{-1}\text{U}_{0}^{-1}(x)\Big] u^{-1}
\end{equation}
We now construct a transformation $u$ satisfying ($\ref{3-18}$) such that $h = {\text{U}^{\prime}}^{u^{-1}}$ 
satisfies the following
 \begin{equation}
\label{4-15}
\begin{aligned}
& h = e^{i A}, \hspace{0.2 cm} |A| < B_{1} (a_{0} + a_{1}),  \hspace{0.1 cm}
|\partial A| < B_{1} (a_{0} + a_{1}), \hspace{0.1 cm} |\Delta A| < B_{1} (a_{0} + a_{1}) \hspace{0.2 cm} \text{on} 
 \hspace{0.1 cm} \Omega_{k}  \\
 & Q_{k}(\text{U}_{0}, A) = B \hspace{0.2 cm} \text{on} \hspace{0.2 cm} \mfB_{k}
 \hspace{0.2 cm} \text{where} \hspace{0.2 cm} B = \frac{1}{i}\text{log}\hspace{0.05 cm}\text{V}(\bar{\text{U}}_{0}^{(k)})^{-1},
 \hspace{0.2 cm} |B| \leqslant 2 a_{1}  
\end{aligned}
\end{equation}
where $Q_{k}(\text{U}_{0}, A)$ is given by .
\begin{equation}
\label{43-16}
Q_{k}(\text{U}_{0}, A) =  \frac{1}{i}\text{log}\hspace{0.05 cm}\overline{h \text{U}_{0}}^{(k)}(\bar{\text{U}}_{0}^{(k)})^{-1}
\end{equation}

\textbf{Theorem 2} 
For arbitrary configurations $\text{U}_{0}$ and $\text{U}^{\prime} \text{U}_{0}$ satisfying $(\ref{3-15})$ 
there exists a constant $c_{1}$ such that if $a_{0} > 0$ and $a_{1} > 0$ obey $a_{0} + a_{1} \leqslant c_{1}$ 
then there is exactly one transformation $u$ satisfying $(\ref{3-18})$ on $\mfB_{k}$ such that the conditions $(\ref{4-15})$ 
hold for the configuration $h = {\text{U}^{\prime}}^{u^{-1}}$.

\textit{Proof} The proof of the Theorem is by induction. First note that the conditions $\text{U}_{0} \in \mfU_{k}(a_{0})$
and $ \text{U}^{\prime} \text{U}_{0} \in \mfU_{k}(a_{0}) \cap \mfB_{k}(\mfB_{k}, \text{V})$ have an inductive property
that they hold for $k-1$ if they are true for $k$. Whereas the condition 
$|(\overline{\text{U}^{\prime} \text{U}_{0}})^{(k)} -  \bar{\text{U}}_{0}^{(k)}| < a_{1}$ on $\mfB_{k}$ does not have the same inductive
property. To apply induction, we follow Balaban and use an equivalent condition (Eq. 1.66,  \cite{B_spaces})
\begin{equation}
\label{5-18}
|(\overline{\text{U}^{\prime} \text{U}_{0}})^{(k)} -  \bar{\text{U}}_{0}^{(k)}|
= |\tilde{\text{U}}^{(k)} - 1| < a_{1}
\end{equation}
where $\tilde{\text{U}}^{(k)} = (\overline{\text{U}^{\prime} \text{U}_{0}})^{(k)}  (\bar{\text{U}}_{0}^{(k)})^{-1} $.
Now we assume that the Theorem is true for $k-1$. This means we have a desired transformation $u_{1}$ 
satisfying $\overline{R_{0}u_{1}}^{(k-1)} = 1$. Then inductively we have
$\overline{R_{0}u_{1}}^{(k)} = 1$ and we construct the desired transformation for $k^{\text{th}}$ step as a perturbation
around $u_{1}$.

Let $u = u^{\prime}u_{1}$ such that $\overline{R_{0}u^{\prime}u_{1}}^{(k)} = 1$. Let $u^{\prime} = e^{i \lambda}$.
Then $u^{\prime}$ is determined by the condition
\begin{equation}
\label{6-18}
\tilde{u} = \overline{R_{0}u^{\prime}u_{1}}^{(k)} (\overline{R_{0}u_{1}}^{(k)})^{-1} = e^{i Q^{\prime}(u_{1}, \lambda)}
= 1 \hspace{0.5 cm} \text{or} \hspace{0.5 cm} Q^{\prime}(u_{1}, \lambda) = 0.
\end{equation}
Then solving $Q^{\prime}(u_{1}, \lambda) = 0$ gives us $\lambda$. Note that we do not need the Landau gauge condition
of Balaban (Eq. 1.80  \cite{B_spaces}). By Proposition 5  \cite{B_spaces}, we get a unique $\lambda$ satisfying
$|\lambda| < \mcO(1) (a_{0} + a_{1})$ for $a_{0} + a_{1} \leqslant c_{1}$.

Let ${\text{U}^{\prime}}^{u_{1}^{-1}} = \text{U}_{1}$. We get the desired configuration as
\begin{equation}
\label{7-18}
h = {\text{U}^{\prime}}^{u_{1}^{-1}{u^{\prime}}^{-1}}= \text{U}_{1}^{{u^{\prime}}^{-1}} = e^{-i\lambda} \text{U}_{1} e^{i \lambda}
\end{equation}
Since $\text{U}_{1}$ is constructed inductively from the solution
of the previous step, we have $\text{U}_{1} = e^{i A^{\prime}}$ with $|A^{\prime}| <  \mcO(1) (a_{0} + a_{1})$. 
Then from the bounds on $A^{\prime}$ and $\lambda$ it follows that 
\begin{equation}
\label{8-18}
\Big|\frac{1}{i} \text{log}\hspace{0.05 cm} {\text{U}^{\prime}}^{u_{1}^{-1}{u^{\prime}}^{-1}} \Big| \leqslant \mcO(a_{0} + a_{1})
\end{equation}
For $k=1$ case take $u_{1} = 1$ such that ${\text{U}^{\prime}}^{u_{1}^{-1}} = \text{U}^{\prime} = \text{U}_{1}$. 
Then from the Lemma 1 \cite{B_spaces}, $|\text{U}^{\prime} - 1| < c \hspace{0.05 cm} a_{1} + a_{2}$ and we then construct
$u^{\prime}$ to get the desired transformation. 

\textit{Uniqueness}. Let two transformations $u_{1}$ and $u_{2}$ satisfy $\overline{R_{0}u_{1}}^{(k-1)} = 1$,
$\overline{R_{0}u_{2}}^{(k-1)} = 1$ respectively. Then by induction they also satisfy 
$\overline{R_{0}u_{1}}^{(k)} = 1$ and $\overline{R_{0}u_{2}}^{(k)} = 1$.
Denote $\text{U}_{1} = {\text{U}^{\prime}}^{u_{1}^{-1}}$ and $\text{U}_{2} = {\text{U}^{\prime}}^{u_{2}^{-1}}$.
Note that
\begin{equation}
\label{9-18}
\overline{R_{0}u_{1}}^{(k)} = \overline{R_{0}u_{1}u_{2}^{-1}u_{2}}^{(k)} = \overline{R_{0}u^{\prime}u_{2}}^{(k)} = 1
\end{equation}
where $u^{\prime} = u_{1}u_{2}^{-1}$. Now by Proposition 5 \cite{B_spaces} there is a unique $u^{\prime}$.
But in the considered case, identity is also the solution since $\overline{R_{0}u_{2}}^{(k)} = 1$.
Therefore, $u^{\prime} = 1$ and $u_{1} = u_{2}$.
This completes the proof of the Theorem 2.

The global transformations $u$ define a one to one map from the space $\mfU_{k}(a_{0}) \cap \mfB_{k}(\mfB_{k}, \text{V})$
into a space of field configurations $h \text{U}_{0}$ with $h$ satisfying the conditions $(\ref{4-15})$.

\textbf{Lemma 3} (\textit{Reduction of proof of Theorem 1}). The variational problem $(\ref{34-21})$ for the functional 
$\mcA^{(k)}(\text{U})$ can be reduced to finding the critical configuration $h$
of the functional $\mcA^{(k)}(h \text{U}_{0}) = \mcA^{(k)}(e^{i A} \text{U}_{0})$ in the space 
\begin{equation}
\label{39-14}
\begin{aligned}
&h = e^{i A}, \hspace{0.2 cm}|A| < \varepsilon_{2}, \hspace{0.2 cm} 
 |\partial A| < \varepsilon_{2}, 
 \hspace{0.2 cm} |\Delta A| < \varepsilon_{2} 
 \hspace{0.1 cm} \text{on} \hspace{0.1 cm} \Omega_{k}, \hspace{0.2 cm} 
 B_{1}(\varepsilon_{0} + \varepsilon_{1}) \leqslant \varepsilon_{2} \\
 & Q_{k}(\text{U}_{0}, A) = B \hspace{0.2 cm} \text{on} \hspace{0.2 cm} \mfB_{k}
 \hspace{0.2 cm} \text{where} \hspace{0.2 cm} B = \frac{1}{i}\text{log}\hspace{0.05 cm}\text{V}(\bar{\text{U}}_{0}^{(k)})^{-1}
 \hspace{0.2 cm} |B| \leqslant 2 \varepsilon_{1} 
\end{aligned}
\end{equation}
where $\text{U}_{0}$ satisfies $(\ref{35-14})$.

\textit{Proof} Let  \underline{$k > 1$}. 
Given a configuration V $ \in \mfB_{k}(\varepsilon_{1})$ so that $\mcV \in \mfB_{k,1}(\varepsilon_{1})$ 
we construct a configuration $\text{V}_{0}$ in the space $\mfB_{k-1}(\varepsilon_{1})$ 
as follows. Take
\begin{equation}
\label{2-13}
\boxed{\text{V}_{0} = \text{V} \hspace{0.2 cm} \text{on} \hspace{0.2 cm} \mfB_{k}  \hspace{0.2 cm} \text{and} \hspace{0.2 cm}
\bar{\text{V}}_{0} = \mcV  \hspace{0.2 cm} \text{on} \hspace{0.2 cm} \mfB_{k,1} \xrightarrow{\text{rescale}}
\text{V}_{0}  \hspace{0.2 cm} \text{on} \hspace{0.2 cm} \mfB_{k-1}}
\end{equation}
We have a configuration $\text{V}_{0}$ as needed for $k-1$ variational problem. 
Assuming that Theorem 1 is true for $k-1$, we get a critical configuration 
$\text{U}_{k-1}(\text{V}_{0}) \in \mfU_{k-1}(B_{3}\varepsilon_{1}) \cap \mfB_{k-1}(\mfB_{k-1}, \text{V}_{0})$.
From the form of regularity conditions and scaling operation in $(\ref{1-19})$ we can say that 
\begin{center}
$\text{U}_{k-1}(\text{V}_{0}) \equiv \text{U}_{0} \in \mfU_{k}(B_{3}\varepsilon_{1})$ 
\end{center}

The actual critical configuration $\text{U}_{k}(\text{V})$ lies in the space
$\mfU_{k}(B_{3}\varepsilon_{1}) \cap \mfB_{k}(\mfB_{k}, \text{V})$. 
We consider the functional $\mcA^{(k)}(\text{U})$ on the space $\mfU_{k}(\varepsilon_{0}) \cap \mfB_{k}(\mfB_{k}, \text{V})$. 
An arbitrary configuration of this space is given by $\text{U} = \text{U}^{\prime} \text{U}_{0}$ as defined in $(\ref{3-15})$.
We take $\text{U}_{0} = \text{U}_{k-1}(\text{V}_{0})$ since $B_{3}\varepsilon_{1} \leqslant \varepsilon_{0}$.
Then by the definition $(\ref{3-15})$ we observe that 
\begin{equation}
\label{35-14}
|\bar{\text{U}}_{0}^{(k)} - \text{V}| < a_{1} \hspace{0.2 cm} \text{on} \hspace{0.2 cm} \mfB_{k}
\end{equation}
This implies that $\text{U}_{0}$ is  close to the critical configuration $\text{U}_{k}(\text{V})$.
Therefore, we use perturbative methods, that is, expand $\mcA^{(k)}(\text{U})$ around $\text{U}_{0}$ 
and find the critical configuration.

To use perturbative methods we make use of the Theorem 2.
Set $a_{0} = \varepsilon_{0}$ and $a_{1} = \varepsilon_{1}$ with
$\varepsilon_{0} + \varepsilon_{1} \leqslant c_{1}$. For $\text{U}_{0} \in \mfU_{k}(\varepsilon_{0})$ and 
$\text{U} = \text{U}^{\prime} \text{U}_{0} \in \mfU_{k}(\varepsilon_{0}) \cap \mfB_{k}(\mfB_{k}, \text{V})$
by Theorem 2 there exists exactly one transformation $u$ satisfying $(\ref{3-18})$
such that $h = {\text{U}^{\prime}}^{u^{-1}}$ satisfies the conditions $(\ref{39-14})$.
 
\underline{$k = 1$}. Note that $k=1$ defines the variational problem at the end of the first iteration.
We take the given configuration V $\in \mfB_{2}(\varepsilon_{1})$ and construct a configuration 
$\text{V}_{0} \in \mfB_{1}(\varepsilon_{1})$ as $(\ref{2-13})$. 
We take $\text{U}_{0} = \text{V}_{0}$ such that $\text{U}_{0} \in \mfU_{1}(\varepsilon_{1})$ and
consider the functional $\mcA^{(1)}(\text{U})$ on the space $\mfU_{1}(\varepsilon_{0}) \cap \mfB_{1}(\mfB_{1}, \text{V}_{0})$
with $\text{U} = \text{U}^{\prime}\text{U}_{0}$ (and  $\text{U}_{0}$ on a larger space  $\mfU_{1}(B_{3}\varepsilon_{1})$).
Then by definition $(\ref{3-15})$ we have $|\bar{\text{U}}_{0} - \text{V}_{0}| < a_{1}$ on $\mfB_{1}$. 
Now we apply Theorem 2 as discussed above.

This completes the proof of Lemma 3.

\textbf{Proposition 4} For $h = e^{i A}$ with variables $A$ in the space $(\ref{39-14})$ there exist variables 
$A_{1}$ and a map $D$ from $A$ to the variables on $\mfB_{k}$ such that the unique critical configuration of the functional  
$\mcA^{(k)}(h \text{U}_{0})$ in the space $|A| < \varepsilon_{2}$ is given by the configuration
\begin{equation}
\label{ }
h = \text{exp}\hspace{0.05 cm} \big[ i \big(A_{1} + H_{1}B - H D(A_{1} + H_{1}B)\big) \big]
\end{equation}
where $|A_{1}| < 2 \varepsilon_{3} $ with $\varepsilon_{2}, \varepsilon_{3}$ sufficiently small and
\begin{equation}
\label{41-14}
4 \varepsilon_{2} \leqslant   \varepsilon_{3} < (18 C_{2} B_{0})^{-1}, 
\hspace{0.5 cm} \varepsilon_{2} < 2 B_{1} \varepsilon_{0},
\hspace{0.5 cm} B_{3}\varepsilon_{1} \leqslant \varepsilon_{0}
\end{equation}
for some constant $C_{2}$.


\subsection{Proof of Proposition 4} \label{expansion}

We start by an expansion of $\mcA^{(k)}(e^{iA} \text{U}_{0})$ about the background $\text{U}_{0}$ as
(Eq. 3.189 \cite{B_prop})
\begin{equation}
\label{42-14}
\mcA^{(k)}(e^{i A} \text{U}_{0}) = \mcA^{(k)}(\text{U}_{0}) + \langle A, \text{J}\rangle +
\frac{1}{2} \langle A, \Delta A\rangle + V_{0}(A)
\end{equation}
where $V_{0}(A)$ denotes $\mcO(A^{3}), \Delta$ is given by (Eq. 3.191 \cite{B_prop}) 
is a small perturbation of the Laplacian that depends on the background and $ \langle A, \text{J}\rangle$ is defined as
\begin{equation}
\label{43-14}
 \langle A, \text{J}\rangle = \sum_{b \subset \T_{\eta}} \eta^{2}\hspace{0.05 cm} \text{Tr} \hspace{0.05 cm}
\big\{-A (\partial b) \eta^{-1} \text{Im}\hspace{0.05 cm} \text{U}_{0}(\partial b) \big\}
\end{equation}
Then we linearize the constraint $Q_{k}$ by a suitable change of variables. We denote the new variables as $A^{\prime}$
and substitute them in $(\ref{42-14})$. After some minor readjustments including another change of variables 
$A^{\prime} \rightarrow A_{1}$ we get a functional whose critical configuration is the one we 
are looking for. We make use of contraction mapping theorem to show the existence of a unique critical configuration.

\subsubsection{Linearizing transformation}
We follow Sect. C of \cite{B_var} on $\Omega_{k}$. The constraint is 
\begin{equation}
\label{32-12}
Q_{k}(\text{U}_{0}, A) = \frac{1}{i}\text{log}\hspace{0.05 cm}(\overline{h\text{U}_{0}}^{(k)})(\bar{\text{U}}_{0}^{(k)})^{-1} = B
\hspace{0.2 cm} \text{on} \hspace{0.2 cm} \mfB_{k}
\end{equation}
It is a non-linear function of $A$. From the definition of the block averaging $(\ref{13-14})$
and using Baker-Campbell-Hausdorff the linear term is $Q_{k} A = \bar{A} + A f(\text{U}_{0})$, where
$f(\text{U}_{0})$ is a function of the background configuration. Then 
\begin{equation}
\label{33-12}
Q_{k}(A) = Q_{k} A + C_{k}(A), \hspace{0.5 cm} |C_{k}(A)| \leqslant C_{2} |A|^{2}
\end{equation}
Here we will  linearize the constraint so that $Q_{k}$ is linear. 

The linearizing transformation is constructed as
\begin{equation}
\label{35-12}
\boxed{A = A^{\prime} - H D(A^{\prime})}
\end{equation}
where $D$ will be a mapping defined on configurations $A$ and taking values in configurations on $\mfB_{k}$.
In fact we will see that $D$ is related to $C_{k}$. 

\textbf{Lemma 5} There exists a constant $0 < \varepsilon_{3} \leqslant (18 C_{2} B_{0})^{-1}$ and a function
$D(A^{\prime}) = \mcO({A^{\prime}}^{2})$ with $|D(A^{\prime})| \leqslant 4 C_{2} ||A^{\prime}||^{2}$ such that
the transformation $(\ref{35-12})$ linearizes the constraint $(\ref{33-12})$.

\textit{Proof} The constraint is linear if
\begin{equation}
\label{36-12}
\begin{aligned}
Q_{k}(A) = Q_{k}(A^{\prime} - H D(A^{\prime}))
&=  Q_{k}A^{\prime} -  Q_{k}H D(A^{\prime}) + C_{k}( A^{\prime} -  H D(A^{\prime})) \\
&=  Q_{k}A^{\prime} \hspace{0.4 cm} \text{on} \hspace{0.1 cm} \T^{0}_{\M+\N-k}
\end{aligned}
\end{equation}
which implies
\begin{equation}
\label{37-12 }
Q_{k}H D(A^{\prime}) = C_{k}(A^{\prime} - H D(A^{\prime}))
\end{equation}
Then $D(A^{\prime})$ is a fixed point of the transformation
\begin{equation}
\label{38-12}
X \rightarrow C_{k}(A^{\prime} -  H X) \hspace{0.4 cm} \text{on} \hspace{0.1 cm} \T^{0}_{\M+\N-k}
\end{equation}
We study this fixed point using contraction mapping theorem. Let
\begin{equation}
\label{39-12}
|A^{\prime}| < \varepsilon_{3} \hspace{0.1 cm} \text{on} \hspace{0.1 cm}\Omega_{k};
\hspace{0.2 cm} X = 0 \hspace{0.1 cm} \text{on} \hspace{0.1 cm} \T^{0}_{\M+\N}; \hspace{0.2 cm}
|X| < \frac{\varepsilon_{3}}{B_{0}} \hspace{0.1 cm} \text{on} \hspace{0.1 cm} \mfB_{k}
\end{equation}
Then the map $(\ref{38-12})$ is contractive if (see the discussion below Eq. 54 in \cite{B_var})
\begin{equation}
\label{45-12}
\varepsilon_{3} \leqslant (18 C_{2} B_{0})^{-1}.
\end{equation}

Identifying $X$ with $D(A^{\prime})$ in $(\ref{39-12})$, we get $|D(A^{\prime})| < B_{0}^{-1}\varepsilon_{3}$. 
Since $\varepsilon_{3}$ was arbitrary, we can take it close to 
\begin{equation}
  ||A^{\prime}|| = \sup_{x \in \Omega_{k}} |A^{\prime}(x)| \hspace{0.2 cm} 
\text{then} \hspace{0.2 cm} |D(A^{\prime})| < B_{0}^{-1} ||A^{\prime}||
\end{equation}
Using $|D(A^{\prime})| < B_{0}^{-1} ||A^{\prime}||$ and from $(\ref{33-12}), (\ref{33-14})$
\begin{equation}
\label{47-12}
|D(A^{\prime})| = |C_{k} (A^{\prime} - H D(A^{\prime}))| \leqslant 
C_{2}( |A^{\prime}| + B_{0}|D(A^{\prime})|)^{2} \leqslant 4 C_{2} ||A^{\prime}||^{2}
\end{equation}
Note that $D(A^{\prime}) = \mcO({A^{\prime}}^{2})$. Let $C_{k}^{(n)}$ and $D^{(n)}$ be homogeneous polynomials of
$n^{\text{th}}$ order, then write
\begin{equation}
\label{48-12}
C_{k}(A^{\prime} - H D(A^{\prime})) = \sum_{n=2}^{\infty} 
C_{k}^{(n)}\big( A^{\prime} - H \sum_{m=2}^{\infty} D^{(m)}(A^{\prime})\big)
= \sum_{n=2}^{\infty} D^{(n)}(A^{\prime})
\end{equation}
and therefore,
\begin{equation}
\label{49-12}
D^{(2)}(A^{\prime}) = C_{k}^{(2)}(A^{\prime}), \hspace{0.2 cm}
D^{(3)}(A^{\prime}) = C_{k}^{(3)}(A^{\prime}) - 2 C_{k}^{(2)}( A^{\prime},  H C^{(2)}(A^{\prime}))
\end{equation}
This completes the proof of the Lemma 5.

\underline{\textit{Locality of $D(A^{\prime})$}}. Note that $D^{(2)}(A^{\prime})$ is ultra localized at a site whereas
$D^{(m)}(A^{\prime})$ for $m > 2$, contains the operator $H$ which has an exponential decaying kernel. Thus, we can say
that functional derivative of $D(A^{\prime})$ decays exponentially. Hence, $D(A)$ is localized.

\textbf{Proposition 6} The transformation $(\ref{35-12})$ satisfying $(\ref{36-12})$ i.e. linearizing $Q_{k}(A)$
is defined and analytic for $A^{\prime}$ satisfying $(\ref{39-12})$ with $\varepsilon_{3} < (18 C_{2} B_{0})^{-1}$;
sufficiently small. The range of this transformation contains the set $(\ref{39-14})$ with 
$\varepsilon_{2} \leqslant \frac{1}{4} \varepsilon_{3}$ and is contained in the corresponding set with $2 \varepsilon_{2}$
instead of $\varepsilon_{3}$. The function $D(A^{\prime})$ satisfies the bound $(\ref{47-12})$ and its functional derivative
decays exponentially.

\textit{Proof}  $D(A^{\prime})$ is a power series in $A^{\prime}$ as $(\ref{48-12})$
and the bound $(\ref{47-12})$ implies that the transformation $A = A^{\prime} - H D(A^{\prime})$ is analytic on
$A^{\prime}$ satisfying $|A^{\prime}| < \varepsilon_{3}$ on $\Omega_{k}$. Then $A$ satisfies
\begin{equation}
\label{50-12}
\begin{aligned}
|A| \leqslant |A^{\prime}| + B_{0} 4 C_{2} ||A^{\prime}||^{2}
&< \varepsilon_{3}  + 4 B_{0} C_{2} \varepsilon_{3}^{2}  \\
&= (\varepsilon_{3} + 4 B_{0} C_{2} \varepsilon_{3}^{2})
\end{aligned}
\end{equation}
since from $(\ref{45-12})$, $\varepsilon_{3} < (4 C_{2} B_{0})^{-1}$ this implies 
\begin{equation}
\label{51-12}
|A| < 2 \varepsilon_{3} \hspace{0.4 cm} \text{on} \hspace{0.1 cm} \Omega_{k}
\end{equation}
Similarly, using $\varepsilon_{3} < (4 C_{2} B_{0})^{-1}$ and $|\partial A^{\prime}| < \varepsilon_{3}$ implies that on $\Omega_{k}$
\begin{equation}
\label{52-12}
|\partial A| \leqslant |\partial A^{\prime}| + B_{0}  4 C_{2} ||A^{\prime}||^{2} 
< 2 \varepsilon_{3} 
\end{equation}
and from $(\ref{45-12})$ that the transformation $A = A^{\prime} - H D(A^{\prime})$ is defined for 
$\varepsilon_{3} \leqslant (18 C_{2} B_{0})^{-1}$ on $\Omega_{k}$.

Now see Eqns. 59-62 in \cite{B_var} in addition to our discussion. This completes the proof of the Proposition 6.

Denote the linearized constraint $Q_{k}(A)$ as $Q$.

\subsubsection{Equations for a solution of the variational problem}

Make change of variables as $A = A^{\prime} - H D(A^{\prime})$ in $(\ref{42-14})$ and consider the functional
\begin{equation}
\label{58-12}
\begin{aligned}
F(A^{\prime}) = \mcA^{(k)}(\text{U}_{0}) &+ \langle A^{\prime} - H D(A^{\prime}), \text{J}\rangle +
\frac{1}{2} \langle A^{\prime} - H D(A^{\prime}), \Delta (A^{\prime} - H D(A^{\prime}))\rangle \\
&+ V_{0}(A^{\prime} - H D(A^{\prime}))
\end{aligned}
\end{equation}
on the space of configurations $A^{\prime}$ satisfying
\begin{equation}
\label{59-12}
Q A^{\prime} = B  \hspace{0.1 cm} \text{on} \hspace{0.1 cm} \T^{0}_{\M+\N-k}, |B| < 2 \varepsilon_{1}; \hspace{0.2 cm}
|A^{\prime}| < \varepsilon_{3}, |\partial A^{\prime}| < \varepsilon_{3}
\hspace{0.1 cm} \text{on} \hspace{0.1 cm} \Omega_{k}
\end{equation}
Note that in the functional $F(A^{\prime})$ the zeroth order and the first order terms in $A^{\prime}$ are 
$\mcA^{(k)}(\text{U}_{0})$ and $\langle A^{\prime}, \text{J}\rangle$ respectively. Next we separate the quadratic terms.
Write
\begin{equation}
\label{60-12}
D(A^{\prime}) = D^{(2)}(A^{\prime}) + D_{3}(A^{\prime});  \hspace{0.2 cm} \text{where} \hspace{0.2 cm}
D^{(2)}(A^{\prime}) = C^{(2)}(A^{\prime}) = C_{k}^{(2)}(A^{\prime}) 
\hspace{0.1 cm} \text{on} \hspace{0.1 cm} \T^{0}_{\M+\N-k}
\end{equation}
Then
\begin{equation}
\label{61-12}
\langle H D(A^{\prime}), \text{J}\rangle = \langle H C^{(2)}(A^{\prime}), \text{J}\rangle + \langle H D_{3} (A^{\prime}), \text{J}\rangle
\end{equation}
and the quadratic form in the expansion of $F(A^{\prime})$ is equal to
\begin{equation}
\label{62-12}
\frac{1}{2} \langle A^{\prime}, \Delta A^{\prime} \rangle - \langle H C^{(2)}(A^{\prime}), \text{J}\rangle 
\equiv \frac{1}{2} \langle A^{\prime}, \Delta_{1} A^{\prime} \rangle 
\end{equation}
Higher order terms determine the functional
\begin{equation}
\label{63-12}
\begin{aligned}
V(A^{\prime}) = - & \langle H D_{3} (A^{\prime}), \text{J}\rangle -  \langle A^{\prime}, \Delta H D(A^{\prime})\rangle \\
&+ \frac{1}{2} \langle H D(A^{\prime}), \Delta H D(A^{\prime})\rangle + V_{0}(A^{\prime} - H D(A^{\prime}))
\end{aligned}
\end{equation}
It is analytic in $A^{\prime}$ for $A^{\prime}$ with values in complex Lie algebra; $\g^{c}$ and satisfying $(\ref{59-12})$.
Rewrite the functional $F(A^{\prime})$ as
\begin{equation}
\label{64-12}
F(A^{\prime}) = \mcA^{(k)}(\text{U}_{0}) + \langle A^{\prime}, \text{J}\rangle + \frac{1}{2} \langle A^{\prime}, \Delta_{1} A^{\prime} \rangle 
+ V(A^{\prime})
\end{equation}
To find the critical points of this functional on the space $(\ref{59-12})$ we have to find $A^{\prime}$ such that
\begin{equation}
\label{65-12}
\Big\langle \frac{\delta}{\delta A^{\prime}} F(A^{\prime}), \delta A^{\prime} \Big\rangle = 0
\end{equation}
for all $\delta A^{\prime}$ in the tangent space to the integration manifold given by
\begin{equation}
\label{66-12}
Q\delta A^{\prime} = 0
\end{equation}
(since the constraint is now linear it lies in the normal space to the integration manifold).

\textbf{Lemma 7} There exists a variable $A_{1}$ with $|A_{1}| < 2 \varepsilon_{3}$ and an operator $\mfG$
with $Q \mfG = 0$ such that the critical equation $(\ref{65-12})$ with constraint $(\ref{66-12})$ is given by $(\ref{78-12})$.

\textit{Proof} 
From $(\ref{64-12})$
\begin{equation}
\label{67-12}
\Big\langle \frac{\delta}{\delta A^{\prime}} F(A^{\prime}), \delta A^{\prime} \Big\rangle =
 \langle \delta A^{\prime}, \text{J}\rangle + \langle \delta A^{\prime}, \Delta_{1} A^{\prime} \rangle + 
 \Big\langle \frac{\delta}{\delta A^{\prime}} V(A^{\prime}), \delta A^{\prime} \Big\rangle
\end{equation}
Thus, we have
\begin{equation}
\label{68-12}
\langle \delta A^{\prime}, \text{J}\rangle + \langle \delta A^{\prime}, \Delta_{1} A^{\prime} \rangle + 
\Big\langle \frac{\delta}{\delta A^{\prime}} V(A^{\prime}), \delta A^{\prime} \Big\rangle = 0 
\hspace{0.1 cm} \text{with} \hspace{0.1 cm} Q \delta A^{\prime} = 0 \hspace{0.1 cm} 
\text{on} \hspace{0.1 cm} \T^{0}_{\M+\N-k}
\end{equation}
As we are working with the operator $\Delta_{1}$ it is better to make a change of variables as
\begin{equation}
\label{70-12}
\boxed{A^{\prime} = A_{1} + H_{1}B}
\end{equation}
and let
\begin{equation}
\label{71-12}
2 B_{1} \varepsilon_{1} < \varepsilon_{3} \hspace{0.2 cm} \text{so that} \hspace{0.2 cm} |H_{1} B| < \varepsilon_{3} \hspace{0.2 cm}
 \text{and} \hspace{0.2 cm} |A_{1}| < 2  \varepsilon_{3} ,  |\partial A_{1}| < 2  \varepsilon_{3}
\end{equation}
Note that since $Q\delta A^{\prime} = 0$, the definition of $H_{1} (\ref{32-14})$ implies that 
$\langle \delta A^{\prime}, \Delta_{1} H_{1} B \rangle = 0$. Then rewrite $(\ref{68-12})$ as
\begin{equation}
\label{72-12}
\langle \delta A^{\prime}, \text{J}\rangle +  \langle \delta A^{\prime}, \Delta_{1} A_{1} \rangle +
\Big\langle \delta A^{\prime}, \Big(\frac{\delta}{\delta A^{\prime}} V\Big) (A_{1} + H_{1}B) \Big\rangle = 0 
\hspace{0.1 cm} \text{with} \hspace{0.1 cm} Q \delta A^{\prime} = 0
 \hspace{0.1 cm} \text{on} \hspace{0.1 cm}\T^{0}_{\M+\N-k}
\end{equation}
The set of configurations $A^{\prime}$ restricted by $(\ref{59-12})$ is contained in the range of the translation 
$A^{\prime} = A_{1} + H_{1}B$ defined on the set of $A_{1}$ restricted by $(\ref{71-12})$. Note that if we take
$A_{1}$ satisfying $(\ref{71-12})$ then the image of the translation is contained in the set of $A^{\prime}$ satisfying
$(\ref{59-12})$ with $3 \varepsilon_{3}$ instead of $\varepsilon_{3}$.

Next we want to get rid off $\delta A^{\prime}$ from $(\ref{72-12})$. We first recall the \textit{projection operator} 
$\mfP$ defined (Eq. 3.139 in \cite{B_prop}) as orthogonal projection onto the subspace of $A^{\prime}$ satisfying the 
condition $Q A^{\prime} = 0$ in a Hilbert space with the scalar product $\langle A^{\prime}, G_{1}^{-1} A^{\prime}\rangle$
where $G_{1}^{-1} = \Delta_{1} + Q^{\ast}aQ$ and  $\Delta_{1}$ is given by $(\ref{62-12})$. 
From Eq. 3.147 in \cite{B_prop} $\mfP$ has a representation
\begin{equation}
\label{-14}
\mfP = I - G_{1} Q^{\ast}(QG_{1}Q^{\ast})^{-1}
\end{equation}
We take $\delta A^{\prime} =  \mfP \delta A$ such that $Q\delta A^{\prime} = 0$ for any arbitrary $A$. Then
\begin{equation}
\label{73-12}
\begin{aligned}
&\langle \delta A^{\prime}, \text{J}\rangle = \langle \mfP\delta A, \text{J}\rangle = \langle \delta A, \mfP^{\ast}\text{J}\rangle \\
&\langle \delta A^{\prime}, \Delta_{1} A_{1} \rangle = \langle \mfP \delta A, (\Delta_{1} + Q^{\ast} a Q) A_{1}\rangle
= \langle \delta A, (\Delta_{1} + Q^{\ast} a Q) \mfP A_{1}\rangle = \langle \delta A, \Delta_{1} A^{\prime} \rangle \\
&\Big\langle \delta A^{\prime}, \Big(\frac{\delta}{\delta A^{\prime}} V\Big) (A_{1} + H_{1}B) \Big\rangle =
\Big\langle \delta A, \mfP^{\ast}\Big(\frac{\delta}{\delta A^{\prime}} V\Big) (A_{1} + H_{1}B) \Big\rangle
\end{aligned}
\end{equation}
where $\delta A$ is now arbitrary. Rewrite $(\ref{72-12})$ as
\begin{equation}
\label{74-12}
\mfP^{\ast}\text{J} + \Delta_{1} A_{1} + \mfP^{\ast}\Big(\frac{\delta}{\delta A^{\prime}} V\Big) (A_{1} + H_{1}B) = 0
\hspace{0.2 cm} \text{with} \hspace{0.2 cm} Q A_{1} = 0
\end{equation}
Denote
\begin{equation}
\label{75-12}
G_{1} = (\Delta_{1} + Q^{\ast} a Q)^{-1} \hspace{0.2 cm} \text{so that} \hspace{0.2 cm}
G_{1}^{-1} A_{1} = \Delta_{1} A_{1}
\end{equation}
and $(\ref{74-12})$ is
\begin{equation}
\label{76-12}
A_{1} + G_{1}\mfP^{\ast}\text{J} + G_{1}\mfP^{\ast}\Big(\frac{\delta}{\delta A^{\prime}} V\Big) (A_{1} + H_{1}B) = 0
\end{equation}
Denote operator
\begin{equation}
\label{77-12}
G_{1}\mfP^{\ast} = \mfG \hspace{0.2 cm} \text{such that} \hspace{0.2 cm} Q\mfG = 0
\end{equation}
and rewrite $(\ref{76-12})$ as
\begin{equation}
\label{78-12}
\boxed{A_{1} + \mfG\text{J} + \mfG \Big(\frac{\delta}{\delta A^{\prime}} V\Big) (A_{1} + H_{1}B) = 0}
\end{equation}
Any solution of $(\ref{76-12})$ or $(\ref{78-12})$ also satisfies $(\ref{74-12})$. $(\ref{78-12})$  is the required 
equation for the solution of the variational problem.

\subsubsection{An analysis of Equation $(\ref{78-12})$}

Let the configuration $A_{1}$ be in the space
\begin{equation}
\label{82-12}
|A_{1}| < \varepsilon_{4}, \hspace{0.1 cm} |\partial A_{1}| < \varepsilon_{4} 
 \hspace{0.1 cm} \text{on} \hspace{0.1 cm} \Omega_{k}
\end{equation}
A solution of Equation $(\ref{78-12})$ is a fixed point of the transformation
\begin{equation}
\label{83-12}
A_{1} \rightarrow - \mfG\text{J} -  \mfG \Big(\frac{\delta}{\delta A^{\prime}} V\Big) (A_{1} + H_{1}B)
\end{equation}
\textbf{Proposition 8} For $\varepsilon_{1}$ satisfying $2 B_{0} B_{3}\varepsilon_{1} \leqslant \varepsilon_{4}$, 
Eq. $(\ref{78-12})$ has exactly one solution in the space $(\ref{82-12})$ which is the fixed point of the transformation
$(\ref{83-12})$. This solution satisfies the bounds $(\ref{82-12})$ with  
$\varepsilon_{4} = 3 B_{0} B_{3}$. 

\textit{Proof} See Sect. E in \cite{B_var}. 

\underline{\textit{To summarize}} On $\Omega_{k}$, from $(\ref{71-12}), |A_{1}| < 2 \varepsilon_{3}$ and from 
$(\ref{82-12}), |A_{1}| <  \varepsilon_{4} $. Since from Proposition 6, $\varepsilon_{3} \geqslant 4 \varepsilon_{2}$
we can take $\varepsilon_{4} = 8 \varepsilon_{2}$ and
$2 B_{0} B_{3}\varepsilon_{1} < B_{1} B_{3}\varepsilon_{1} \leqslant \varepsilon_{2}$. Then Proposition 8 is true.
Also note that from $(\ref{39-14})$ we have $B_{1}(\varepsilon_{0} + \varepsilon_{1}) 
\leqslant \varepsilon_{2}$. Let us then take $\varepsilon_{2} = B_{1} \varepsilon_{0} + B_{1} B_{3}\varepsilon_{1}$
so that this condition is satisfied automatically. Now using $B_{3}\varepsilon_{1} \leqslant \varepsilon_{0}$ we have
\begin{equation}
\label{90-12}
8 \varepsilon_{2} \leqslant 8 B_{1}\varepsilon_{0} + 8 B_{1} \varepsilon_{0} \leqslant 16 B_{1} \varepsilon_{0}
\end{equation}
This completes the proof of Proposition 4.

\subsection{Proof of Theorem 1} \label{proof1} The proof of Theorem 1 is by induction. 
For $k=1$, a fixed configuration $\text{V}_{0} \in \mfB_{1}(\varepsilon_{1})$ is given. We define $\text{U}_{0} = \text{V}_{0}$
such that $\text{U}_{0} \in \mfU_{1}(\varepsilon_{1})$. Then we consider the functional $\mcA^{1}(\text{U})$
on the space $\mfU_{1}(\varepsilon_{0}) \cap  \mfB_{1}(\mfB_{1}, \text{V}_{0})$. 
The definition $(\ref{3-15})$ gives us freedom to consider $\text{U}_{0}$ on a larger space bounded by $\varepsilon_{0}$.
Thus, we can have a large constant $B_{3}$ obeying $B_{3}\varepsilon_{1} \leqslant \varepsilon_{0}$ and take
$\text{U}_{0} \in \mfU_{1}(B_{3}\varepsilon_{1})$ such that for $B_{3} = 1$, we have $\bar{\text{U}}_{0} = \text{V}_{0}$
on $\mfB_{1}$.

Now take $\text{U} = \text{U}^{\prime}\text{U}_{0}$. 
Next we make use of the mapping transformation (Theorem 2) and get the functional
$\mcA^{1}(h\text{U}_{0}) = \mcA^{1}(e^{iA}\text{U}_{0})$ to solve the critical problem perturbatively as shown in Lemma 3. 
Then Proposition 4 gives us a  unique critical configuration of  $\mcA^{1}(e^{iA}\text{U}_{0})$ in the space 
$|A| < \varepsilon_{2} < \mcO(1)\varepsilon_{0}$.

To understand what $B_{3}$ is like let $\Box$ denote a $M-$ block and $\tilde{\Box}$ denote a block of linear
size $3 M$ consisting of nearest neighbors of $\Box$. Denote $\Box^{\prime} \in \Box^{c} \cap \tilde{\Box}$.
Let $y_{1} \in \Box$. We follow Sect. F (Eqs. 162, 163) in \cite{B_var} and write
\begin{equation}
\label{ }
B_{3} = \mcO(1) B_{0} \sup_{\Box^{\prime}} \sup_{y_{1}} \sum_{y_{2} \in \Box^{\prime}} e^{-\kappa d(y_{1}, y_{2})} (d(y_{1}, y_{2}) + 1)
\end{equation}
and assume that $B_{3} \hspace{0.05 cm} e^{-\kappa (2M)} \leqslant \frac{1}{2}$. This completes the proof for $k=1$.

For each inductive step $k > 1$, the critical configuration for $k-1$ serves as $\text{U}_{0} \in \mfU_{k}(B_{3}\varepsilon_{1})$
for the $k^{\text{th}}$ step,
where $B_{3}$ is provided by the inductive hypothesis on $k-1$. Then we consider the functional  $\mcA^{k}(\text{U})$
on the space $\mfU_{k}(\varepsilon_{0}) \cap  \mfB_{k}(\mfB_{k}, \text{V})$ and solve the critical problem perturbatively 
as in Lemma 3 and Proposition 4.

\subsubsection{Critical configuration is a minimum}

Let $\text{U}_{k} = h \text{U}_{0} \in \mfU_{k}(B_{3}\varepsilon_{1})  \cap \mfB_{k}(\mfB_{k}, \text{V})$.

To see that $\text{U}_{k}$ is a minimum we apply the whole procedure again with $\text{U}_{k}$
instead of $\text{U}_{0}$. We get a functional $F(A^{\prime})$ as $(\ref{64-12})$ for which the critical configuration
is $A^{\prime} = 0$. Then from $(\ref{65-12})$, $(\ref{66-12})$, $(\ref{67-12})$ together with $A^{\prime} = 0$ we conclude that
\begin{equation}
\label{95-12}
\langle \delta A^{\prime}, \text{J} \rangle = 0 \hspace{0.2 cm} \text{for all} \hspace{0.2 cm}
\delta A^{\prime} : Q\delta A^{\prime} = 0 
\end{equation}
But due to the shift to $\text{U}_{k}$ we also have $QA=0$ (the constraint is automatically satisfied
since $\bar{\text{U}}_{k}^{(k)} = \text{V}$). Hence the configuration $A^{\prime}$ satisfies the same conditions as $\delta A^{\prime}$
and we have $\langle A^{\prime}, \text{J} \rangle = 0$. Thus,
\begin{equation}
\label{96-12}
F(A^{\prime}) = \mcA^{(k)}(\text{U}_{k}) + \frac{1}{2} \langle A^{\prime}, \Delta_{1} A^{\prime} \rangle + V(A^{\prime})
\end{equation}
since $\Delta_{1}$ is positive definite, the second order differential at $A^{\prime} = 0$ is positive. Hence,
$A^{\prime} = 0$ is a minimum. Thus, $\text{U}_{k}$ is a minimal configuration of the functional $\mcA^{(k)}(\text{U})$.

This completes the proof of Theorem 1.

\section{Analyticity of the Background configuration}\label{Analyticity}

We have mentioned before that the fluctuation about the background configuration
are the integration variables in a renormalization group analysis. 
Here we want to conclude that the minimal configuration
we get as a result of the variational problem is an analytical function of the fluctuation variables.
 
The minimal configuration $\text{U}_{k}$ is $\text{U}_{k}(\text{V})$. 
Let $\text{V} = \text{V}^{\prime}\text{V}_{0}$ with $\text{V}^{\prime}$ small
and $\text{U}_{0} = \text{U}_{k}(\text{V}_{0})$. Denote the fluctuation variable as 
$B = \frac{1}{i} \text{log}\hspace{0.05 cm} \text{V}^{\prime}$.

Now take $\text{V} = \text{V}^{\prime}\text{V}_{0}$ with $|\text{V}^{\prime} - 1| < \varepsilon_{1}$ on $\mfB_{k}$.
Then
\begin{equation}
\label{97-12}
\text{U}_{k}(\text{V}) = \text{U}_{k}(e^{iB}\text{V}_{0}) = e^{i \mcH(B)} \text{U}_{0}
\end{equation}
where $|\mcH(B)| < \varepsilon_{2}$ as $(\ref{39-14})$ on $\Omega_{k}$.
Then from Proposition 3, the configuration $\mcH(B)$ has a representation 
\begin{equation}
\label{98-12}
\mcH = A_{1} + H_{1}B - H D(A_{1} + H_{1}B)
\end{equation}
where $A_{1}$ satisfies
\begin{equation}
\label{99-12}
A_{1} + \mfG\Big(\frac{\delta}{\delta A^{\prime}} V \Big) (A_{1} + H_{1}B) = 0
\end{equation}
with $\varepsilon_{2}$ satisfying $(\ref{41-14})$. $A_{1}$ as a solution of $(\ref{99-12})$ is an analytic
function of $H_{1}B$. Since $V$ starts with $V^{(3)}$; the terms of the third order, $A_{1}$ starts with
second order in $H_{1}B$. Then from Eqs. 196, 197 in \cite{B_var}
\begin{equation}
\label{100-12}
\begin{aligned}
\mcH &= H_{1}B -  \mfG\Big(\frac{\delta}{\delta A^{\prime}} V^{(3)} \Big)(H_{1}B) - H C^{(2)}(H_{1}B) + \cdots \\
\mcH &= H_{1}B + \mcO(B^{2})
\end{aligned}
\end{equation}
where we have used $(\ref{49-12})$ for writing $D$ in terms of $C^{(2)}$.

\subsection{Invariance of the minimal configuration}
We have shown that the minimal configuration $\text{U}_{k}(\text{V}) = e^{i \mcH(B)} \text{U}_{0}$ is analytic in the domain 
$|\mcH(B)| < \varepsilon_{2}$. Under a symmetry transformation $u\text{V}u^{-1}$, $\text{U}_{k}(u\text{V}u^{-1})$
is defined as long as it is analytic. For $u \in \mcG(\text{V})$, we have
\begin{equation}
\label{ }
\text{U}_{k}(u\text{V}u^{-1}) = \text{U}_{k}(\text{V}).
\end{equation}

\textbf{Acknowledgement} The work is supported by the National Science Center (NCN), Poland 
Grant number: 2019/34/E/ST1/00053.


\begin{thebibliography}{99}
 
  \bibitem{B_6}  
 {Balaban, T.}
 {Propagators and renormalization transformations for lattice gauge field theories-I},
  {Commun. Math. Phys.} {95: 17-40}, {(1984)}.
  
   \bibitem{B_7}  
 {Balaban, T.}
 {Propagators and renormalization transformations for lattice gauge field theories-II},
  {Commun. Math. Phys.} {96: 223-250}, {(1984)}.
  
   \bibitem{B_avg}  
 {Balaban, T.}
 {Averaging Operations for Lattice Gauge Theories},
  {Commun. Math. Phys.} {98: 17-51}, {(1985)}.
  
  
 \bibitem{B_prop}  
 {Balaban, T.}
 {Propagators for Lattice Gauge Theories in a Background Field},
  {Commun. Math. Phys.} {99, 389-434}, {(1985)}.
  
   \bibitem{B_10}  
 {Balaban, T.}
 {Ultraviolet stability of three-dimensional lattice pure gauge field theories},
  {Commun. Math. Phys.} {102: 255-275}, {(1985)}.
  
  \bibitem{B_spaces}  
 {Balaban, T.}
 {Spaces of Regular Gauge Field Configurations on a Lattice and Gauge Fixing Conditions},
 {Commun. Math. Phys.} {99: 75-102}, {(1985)}.

   \bibitem{B_var}  
 {Balaban, T.}
 {The Variational Problem and Background Fields in Renormalization Group Method for Lattice Gauge Theories},
  {Commun. Math. Phys.} {102: 277-309}, {(1985)}.
  
  \bibitem{B_12}  
 {Balaban, T.}
 {Renormalization group approach to lattice gauge field theories-I},
  {Commun. Math. Phys.} {109: 249-301}, {(1987)}.
  
 \bibitem{B_13}  
 {Balaban, T.}
 {Renormalization group approach to lattice gauge field theories-II},
  {Commun. Math. Phys.} {116: 1-22}, {(1988)}.
  
   \bibitem{B_14}  
 {Balaban, T.}
 {Convergent renormalization expansions for lattice gauge field theories},
  {Commun. Math. Phys.} {119, 243-285}, {(1988)}.
  
   \bibitem{B_15}  
 {Balaban, T.}
 {Large field renormalization-I},
  {Commun. Math. Phys.} {122:, 175-202}, {(1989)}.
  
   \bibitem{B_16}  
 {Balaban, T.}
 {Large field renormalization-II},
  {Commun. Math. Phys.} {122:, 355-392}, {(1989)}.
  
\bibitem{JW}
{Jaffe, A. and Witten, E.}, {Quantum Yang Mills theory},
{In J. Carlson, A. Jaffe and A. Wiles (Eds.)}, {The millennium prize problems}, {129-152},
{American Mathematical Society}, {(2006)}.

\bibitem{MRS}
{ Magnen, J.; Rivasseau, V. and S$\acute{e}$n$\acute{e}$or, R.},
{Construction of $YM_{4}$ with an Infrared Cutoff},
{Commun. Math. Phys. 155: 325-383 (1993)}. 

 \bibitem{Ab}  
 {Abbott, L. F.}
 {Introduction to the background field method},
  {Acta Physica Polonica}, {Vol. B13}, {(1982)}.
  
 \bibitem{HPS}  
 {Howe, P. S.; Papadopoulos, G. and Stelle, K. S.}
 {The background field method and the non-linear $\sigma$ Model},
  {Nucl. Phys. B} {296: 26-48}, {(1988)}.
  
  \bibitem{D} 
  {Dimock, J.}, {The Renormalization Group according to Balaban - I. Small Fields}
 {Rev. Math. Phys}, {25}, {1330010}, {(2013)},
 
\bibitem{BJ}
{Balban, T. and Jaffe, A.}, {Constructive Gauge Theory},
{In G. Velo and A. S. Wightman (Eds.)}, {Fundamental Problems of Gauge Field Theory}, {207-263},
{Springer}, {(1986)}


\end{thebibliography}
\end{document}